\documentclass[proof]{pasj01}
\draft
\usepackage{natbib, aas_macros}
\usepackage{lscape}
\usepackage{graphicx}
\usepackage{cases}
\Accepted{$\langle$September 27, 2018$\rangle$}
\usepackage{mediabb}
\begin{document}

\title{High entropy and evidence for a merger in the low surface brightness cluster Abell 2399}
\author{Ikuyuki \textsc{Mitsuishi}\altaffilmark{1,}$^{*}$, Yasunori \textsc{Babazaki}\altaffilmark{1}, Naomi \textsc{Ota}\altaffilmark{2}, Shin \textsc{Sasaki}\altaffilmark{3}, Hans \textsc{{B{\"o}hringer}}\altaffilmark{4,5}, Gayoung \textsc{Chon}\altaffilmark{4}, and Gabriel W. \textsc{Pratt}\altaffilmark{6,7}} %

\altaffiltext{1}{Department of Physics, Nagoya University, Furo-cho, Chikusa-ku, Nagoya, Aichi 464-8602, Japan}
\altaffiltext{2}{Department of Physics, Nara Women's University, Kitauoyanishi-machi, Nara, Nara 630-8506, Japan}
\altaffiltext{3}{Department of Physics, Tokyo Metropolitan University, 1-1 Minami-Osawa, Hachioji, Tokyo 192-0397, Japan}
\altaffiltext{4}{Universit\"{a}ts-Sternwarte, Ludwig-Maximilians Universit\"{a}t M\"{u}nchen, Scheinerstr. 1, 81679 M\"{u}nchen, Germany}
\altaffiltext{5}{Max-Planck-Institut f{\"u}r extraterrestrische Physik, D-85748 Garching, Germany}
\altaffiltext{6}{IRFU, CEA, Universit{\'e} Paris-Saclay, F-91191 Gif-sur-Yvette, France}
\altaffiltext{7}{Universit{\'e} Paris Diderot, AIM, Sorbonne Paris Cit{\'e}, CEA, CNRS, F-91191 Gif-sur-Yvette, France}
\email{mitsuisi@u.phys.nagoya-u.ac.jp}


\maketitle

\begin{abstract}
  We present results of the X-ray analyses of the nearby cluster of
  galaxies Abell~2399 ($z=0.058$) using the XMM-Newton and Suzaku
  satellites. This cluster is classified as a low surface brightness (LSB) cluster.  It has a bimodal structure
  in X-rays, and the X-ray-emission peaks are significantly offset
  from the peaks in gas temperature.  By de-projecting the 
  annular spectra, the temperature and electron density in the central $r<0.1r_{200}$ region are measured 
  to be 3.6~keV and $1.4\times10^{-3}~{\rm cm^{-3}}$, respectively.
  %for $r<140$~kpc. 
  This yields a very high gas entropy, $\sim 300~{\rm
    keV\,cm^{2}}$, in the central region, comparable to the values observed
  in other LSB clusters. 
%  at the center
  The scaled entropy of Abell~2399 is the highest among the REXCESS cluster sample. 
  The spatial distribution of the member galaxies
  exhibits multiple substructures, the locations of which are significantly
  different from those in the X-ray image.  We discovered a clear
  discontinuity in the X-ray brightness and temperature distributions
  in the western gas clump, which provides evidence of the
  presence of a cold front.  Therefore, our results strongly suggest that
  A2399 has experienced a merger and that the high central entropy originates from the merger activity.
\end{abstract}

\section{Introduction}
A low surface brightness cluster of galaxies (hereafter an LSB cluster) is
a type of cluster that has an extremely low X-ray surface
brightness and a highly irregular morphology. Several LSB clusters
have been identified in the REXCESS cluster sample, which make up 
5--10\% of the cluster sample detected by the ROSAT All-Sky Survey
\citep{2013A&A...555A..30B,2017AJ....153..220B}.  
One requires deep observations to investigate the detailed gas properties of such clusters 
because of their extremely low brightness (typically $\sim 10^{-14}~{\rm
  erg\,s^{-1}cm^{-2}arcmin^{-2}}$ in the 0.1--2.4~keV band) and
diffuse gas distribution without a prominent core. Hence their evolution remains a puzzle. Since upcoming cluster
surveys--by, e.g., eROSITA on the SRG satellite--are expected to detect a large number
of clusters including faint ones \citep{2012arXiv1209.3114M}, it is
important to understand the properties of LSB clusters, which may not
follow the mass-scaling relations of bright clusters.

The Suzaku
satellite \citep{2007PASJ...59S...1M} has been used to analyze a few LSB clusters in detail. 
A76 is an archetype LSB
cluster; it has the lowest surface brightness among the ROSAT clusters
studied by \cite{1999A&A...348..711N} and an irregular
morphology. \cite{2013A&A...556A..21O} reported that A76 has an
exceptionally high gas entropy in the central region, $K \equiv kT n_{\rm
  e}^{-2/3}\sim400~{\rm keV\,cm^2}$, and a low
electron density, of the order of $10^{-4}-10^{-3}~{\rm cm^{-3}}$, for the
observed mean gas temperature of 3~keV. This result is not explained
either by gravitational heating or by preheating. A similar trend has
been found in other LSB clusters, such as A548W \citep{2016PASJ...68S..21N} and
A1631 \citep{2018PASJ...70...46B}. \cite{2018PASJ...70...46B} noted
that the central entropy of A1631 is one of the highest ($\gtrsim
400~{\rm keV\,cm^2}$) among known nearby clusters, while the X-ray
luminosity is lower by a factor of 3 than that expected from the
luminosity-temperature relation for relaxed clusters. In addition, they found
that the spatial distribution of the gas is significantly different from
that of the member galaxies and they proposed a post-merger scenario to explain
the production of high entropy gas in this irregular cluster.

To clarify the nature of LSB clusters, we need to increase the number of clusters 
in the sample.  In this paper, we conduct an X-ray imaging and spectroscopic
study of the LSB cluster Abell 2399 at $z=0.058$  using the XMM-Newton
and Suzaku satellites. The good spatial resolution of XMM-Newton
enables us to derive the temperature distribution of the hot gas
associated with substructures in the cluster. In addition, we have searched for discontinuities in the X-ray surface brightness as an indicator of the
merger activity.  Suzaku's low-background level allows us to
measure the gas-entropy profile out to the virial radius.  We have also compared the spatial distributions of
the galaxies and the hot gas in this system because 
rich spectroscopic data in optical are available for the member galaxies 
\citep{2017A&A...599A..81M}.
In previous studies, A2399 (the alternative name is RXC~J2157-0747) has been analyzed within the REXCESS sample \citep{2007A&A...469..363B}; the profiles of gas density and entropy were determined up to $\sim$$r_{500}$ by \citet{2008A&A...487..431C} and \citet{2010A&A...511A..85P} and the morphological investigation was also studied by \citet{2010A&A...514A..32B}. They noted that RXC~J2157-0747 has a higher entropy and is one of outliers in the substructure characterization such as a power ratio and a centroid shift parameter among the REXCESS sample even though they did not pursue the origin of the specific properties.

This paper is organized as follows: In section~2 we present the X-ray
observations of A2399 with Suzaku and XMM-Newton and the data
reduction. Sections~3--4 describe the results from our imaging and
spectral analyses.  In section~5, we discuss the X-ray and optical
properties and propose a scenario to interpret the data.  Where
necessary, we assume the standard cosmological model, with a matter
density ${\rm \Omega}_M~=~0.27$, a cosmological constant ${\rm
  \Omega}_\Lambda~=~0.73$, and a Hubble constant $H_0~=~70~{\rm
  km~s^{-1}~Mpc^{-1}}$. At the cluster redshift $z=0.058$, $1\arcmin$
corresponds to 68~kpc. In this paper, we have employed HEAsoft v6.21 and XSPEC
version 12.9 to perform $\chi^2$ fitting of the X-ray spectra, and we have used the metal-abundance table
of \citet{1989GeCoA..53..197A}. Unless otherwise stated, the error
ranges show the 90\% confidence levels from the central values.

\section{Observation and data reduction}

\subsection{Suzaku}\label{sec:reductionsuzaku}
In November 2014, four-pointed Suzaku observations of A2399 (the center, east, west, and
north) were performed.  
Table~\ref{tab:obssummary} summarizes the basic information about the observations. 
Four X-ray sensitive CCD cameras--known as the XIS instruments--are installed on Suzaku. Three of the CCD cameras are front-illuminated (XIS-0,
-2, -3), and the other is back-illuminated (XIS-1) \citep{2007PASJ...59S..23K}. 
During the observation periods, the XIS-0, -1 and -3 CCD cameras were operated in normal mode with 
space charge injection enabled \citep{2009PASJ...61S...9U}.

%%%%%%%%%%%%%%%%%%%%%%%%%%%%%%%%%%%%%%%%%%
\begin{table*}[htb]
\tbl{Logs of Observations of A2399 with Suzaku and XMM-Newton}{%
\begin{tabular}{llllll}
\hline \noalign{\vskip2pt} 
Target               & Obs ID     & Date                &  RA.     &   Dec.   & Net Exposure  \\
                     &            &                     & [deg]    &  [deg]   &  [ks]     \\ \hline
Suzaku/XIS           &            &                     &          &          &       \\              
A2399 West           & 809020010 & 2014-11-11 & 329.1831 & -7.8643 & 23.1 \\ 
A2399 Center         & 809022010 & 2014-11-12 & 329.2851  & -7.5651 & 26.0 \\ 
A2399 North          & 809021010 & 2014-11-12 & 329.3708 & -7.7975 & 18.6 \\ 
A2399 East           & 809023010 & 2014-11-15 & 329.5573 & -7.7299  & 20.9 \\ \hline 
XMM-Newton/EPIC      &         &   &                     &          &              \\
RXC~J2157-0747 (A2399)    & 0654440101 & 2010-06-07 &  329.3573 & -7.7987  &  54.9\footnotemark[$*$] (MOS1), 59.5\footnotemark[$*$] (MOS2), 26.2\footnotemark[$\*$] (PN)   \\ \hline
\end{tabular}
}\label{tab:obssummary}
\begin{tabnote}
\footnotemark[$*$] Net exposure time after removal of periods of high background flaring.
\end{tabnote}
\end{table*}
%%%%%%%%%%%%%%%%%%%%%%%%%%%%%%%%%%%%%%%%%%

%%%%%%%%%%%%%%%%%%%%%%%%%%%%%%%%%%%%%%%%%%
\begin{figure*}[htb]
 \begin{minipage}{0.5\hsize}
 \begin{center}
    \includegraphics[width=80mm,angle=0]{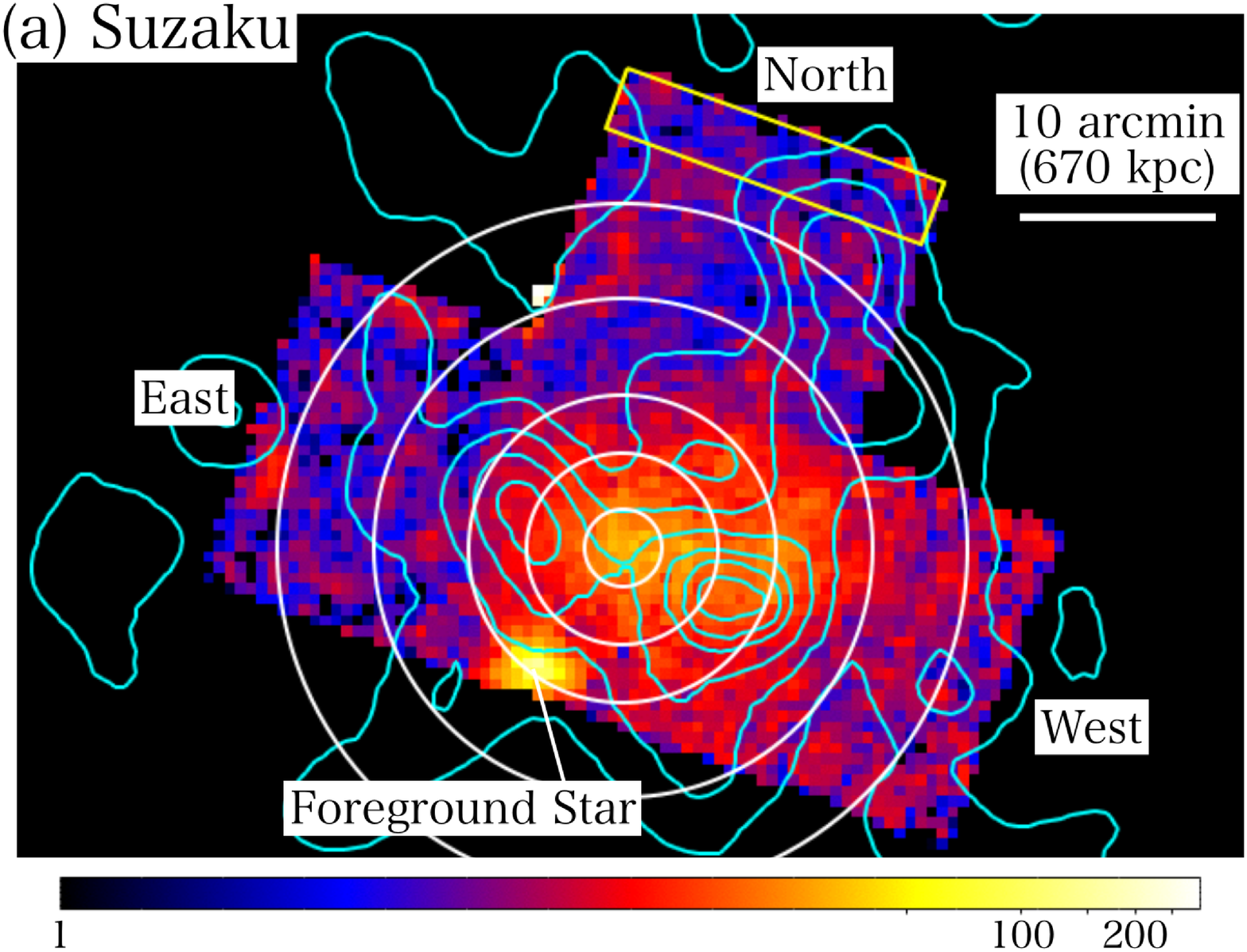}
 \end{center}
 \end{minipage}
 \begin{minipage}{0.5\hsize}
 \begin{center}
    \includegraphics[width=80mm,angle=0]{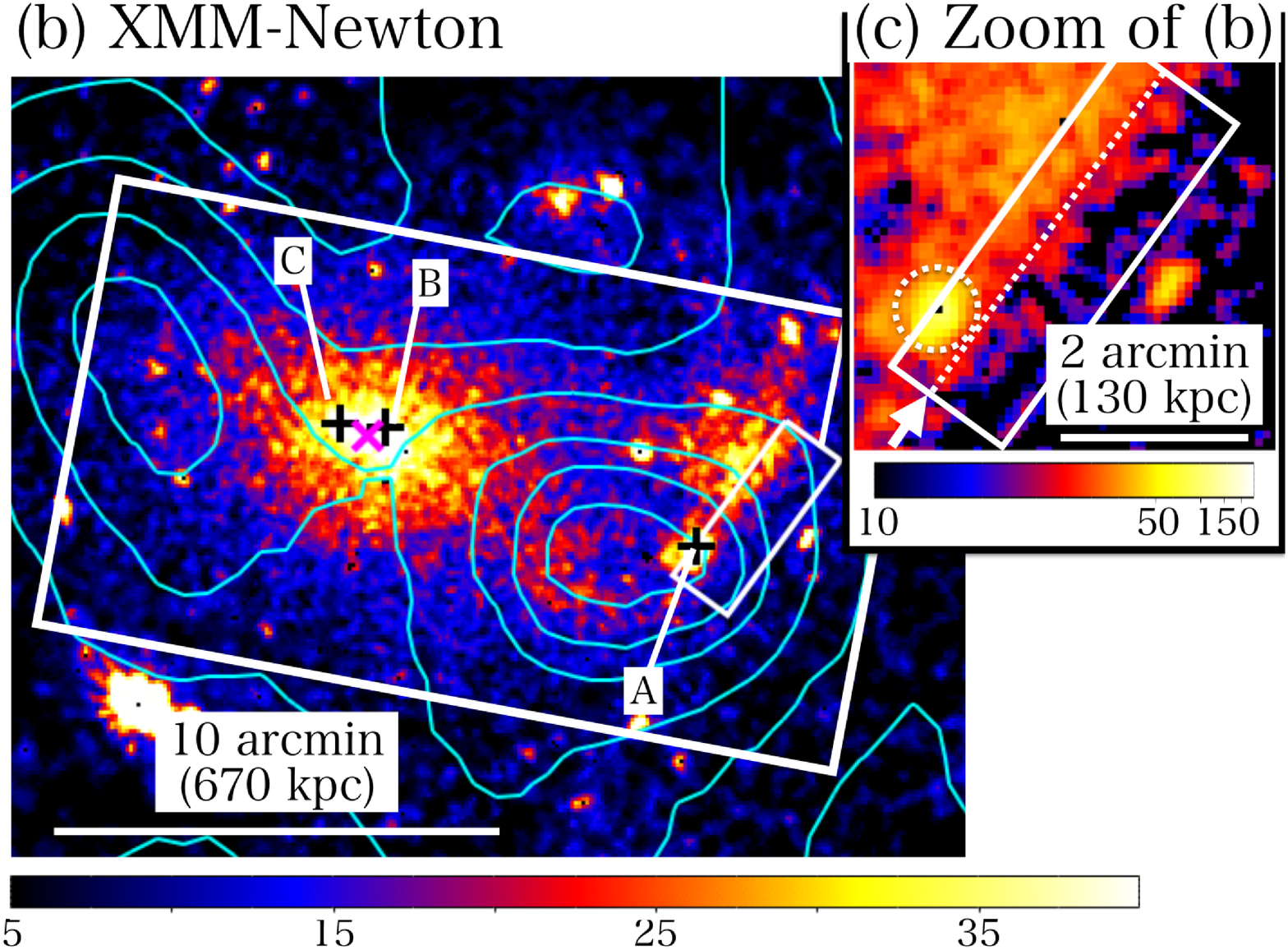}
 \end{center}
 \end{minipage}
 \vspace{0.1cm}
  \caption{Suzaku and XMM-Newton images in the 0.5--5~keV band. Solid white rings and rectangles and cyan contours indicate the regions used in the imaging and spectral analysis and the galaxy number density distribution, respectively. (a) Suzaku XIS-1 mosaic X-ray image of A2399 in units of cts (20 ks)$^{-1}$ (1024
    pixels)$^{-1}$. We subtracted the NXB component. We corrected for exposure and
    vignetting effect and smoothed the image using a
    Gaussian function with $\sigma = 25\arcsec$.
    We used the rectangular region outlined in yellow to estimate the X-ray background
    emission. A bright foreground star, SDSS J215751.40-075348.1, is shown and excluded in the spectral analysis. (b) XMM-Newton EPIC composite
    images in units of cts s$^{-1}$ (degree)$^{-2}$. The QBP background has been subtracted. The positions of the 1st, 2nd,
    and 3rd brightest galaxies are denoted by A, B, and C with black cross marks. The magenta cross shows the X-ray
    peak, which is derived from the XMM-Newton image after excluding point sources. (c) A close-up view of (b) to emphasize an extracted area (a white rectangle) for the cold front study. An arrow and a dotted line show a plane of the discontinuity in the surface brightness. A dotted white circle corresponds to a point source excluded in the analysis.}
\label{fig:a2399img}
\end{figure*}
%%%%%%%%%%%%%%%%%%%%%%%%%%%%%%%%%%%%%%%%%%

We used standard Suzaku pipeline processing to create event files 
with CALDB [v2016-06-07 for XIS and v2011-06-30 for the X-ray telescopes \citep[XRTs:][]{ 2007PASJ...59S...9S}].  
We used the following criteria to screen the data: Earth elevation angle $>5^\circ$,
day-Earth elevation angle $>20^\circ$, and time removal during the
South Atlantic Anomaly.  As an additional correction, we removed flickering
pixels\footnote{https://heasarc.gsfc.nasa.gov/docs/suzaku/analysis/xisnxbnew.html}.
We also excluded point sources detected by XMM-Newton (see subsection~\ref{subsec:xmm}) from this Suzaku analysis.

We determined the X-ray background from the XIS spectra in the
northern field, $\sim1.3$~Mpc away from the center
(figure~\ref{fig:a2399img} (a)).  We subtracted the detector background
and evaluated other background components assuming the same models  as used in, e.g., \citet{2012PASJ...64...18M} and \citet{2018PASJ...70...46B}, which 
consists of the comic X-ray background (CXB) and the Galactic X-ray
background (GXB). Table~\ref{tab:a2399background} summarizes the best-fit
parameters, and they are consistent with typical values for the CXB and GXB components
\citep[e.g.,][]{2002PASJ...54..327K,2009PASJ...61..805Y}. Thus, we
incorporate this background model in the spectral analysis.

%%%%%%%%%%%%%%%%%%%%%%%%%%%%%%%%%%%%%%%%%%
\begin{table*}[htb]
\caption{Best fit parameters for the X-ray background emission}{%
\begin{tabular}{lllllll}
\hline \noalign{\vskip2pt} 
$kT_{\rm LHB}$ & Norm$_{\rm LHB}$\footnotemark[$*$] & $kT_{\rm GH}$ & Norm$_{\rm GH}$\footnotemark[$*$] & $\Gamma_{\rm CXB}$  & Norm$_{\rm CXB}$\footnotemark[$\dagger$] & $\chi^2$/d.o.f \\
keV    &      & keV &  &  &   &  \\ \hline
0.1 (fixed) &  $11.3^{+4.3}_{-7.8}$  &  0.30$^{+0.50}_{-0.12}$ & $5.4^{+1.9}_{-3.5}$ & 1.4 (fixed) & $10.4^{+1.7}_{-1.5}$ & 72/62  \\ \hline
\end{tabular}\label{tab:a2399background}
\begin{tabnote}
\hspace{-1.0cm}\footnotemark[$*$] Normalization of the APEC model divided by a solid angle $\Omega$ (sr). $Norm = (10^{-14}/\Omega)$ $\int{n_{\rm e}n_{\rm H} dV/[4((1 + z) D_A)^2 ]}$ in units of $\ {\rm cm}^{-5} {\rm sr}^{-1}$, \\
\hspace{-0.8cm}where, $n_{\rm e}$, $n_{\rm H}$, $D$, and $V$ are the electron and hydrogen number densities (cm$^{-3}$), the angular diameter distance (cm) and the emission volume (cm$^3$), \\
\hspace{-0.8cm}respectively. Here, ``LHB'' denotes the local hot bubble, and ``GH'' the Galactic halo, respectively.\unskip\\
\hspace{-1.0cm}\footnotemark[$\dagger$] The unit is photons keV$^{-1}$ cm$^{-2}$ s$^{-1}$ sr$^{-1}$ at 1 keV. \unskip\\
\end{tabnote}
}
\end{table*}
%%%%%%%%%%%%%%%%%%%%%%%%%%%%%%%%%%%%%%%%%%

\subsection{XMM-Newton}\label{subsec:xmm}
We retrieved three datasets containing A2399 (RXC~J2157-0747) from the
XMM-Newton Science Archive.  We used only the one data set OBSID
0654440101 in the analysis because two of them are heavily
contaminated by soft-proton flares, .

We performed data reduction for the set of three X-ray CCD cameras on board XMM-Newton--
the European Photon Imaging Camera (EPIC)--in the standard manner, using the ESAS (Extended Source Analysis Software) package
\citep{2008A&A...478..615S} in SAS version 16.0.0.
We removed high background periods, for which the rates were beyond the 2$\sigma$ range
of the rate distribution. 
Table~\ref{tab:obssummary} shows the observation identifications and the net exposure times after filtering. 
We used an ESAS routine to detect point sources in each detector and removed them from our analysis.

\section{X-ray images}\label{sec:xray_image}
Figures~\ref{fig:a2399img} (a) and (b) show Suzaku/XIS-1 and
XMM-Newton/EPIC images of A2399 in the 0.5--5.0~keV band,
respectively.  In the XIS image, diffuse cluster emission is
detected out to the virial radius $r_{200} = 1.2$~Mpc or $17\arcmin$.
Here the virial radius is estimated from the global temperature $\langle kT\rangle=2.6$~keV and the $T-r_{200}$ relation
\citep{2005A&A...441..893A} which is obtained from clusters in the temperature range of 2--9 keV.

In the EPIC image, the cluster shows a bimodal structure; the main gas clump in the east
and a smaller (``the sub'') gas clump in the west each have a projected distance of
500~kpc.  The X-ray surface brightness of the main clump has a peak (329.3790, -7.7982) after removal of point-like sources, and it appears to be extended in the
southwest-northeast direction.  On the other hand, the subclump has a
rather elongated shape, extending in the southeast-northwest direction
\citep[see also ][]{2007A&A...469..363B}.

The bright X-ray point-like sources located around the center of the main / sub
clumps correspond to the three brightest cluster galaxies \citep[BCGs;][]{2004AJ....128.1558S,2009A&A...495..707C,2009A&A...497..667V,2017A&A...599A..81M} around redshift of A2399, i.e., 
WINGS~J215701.72-075022.0, WINGS~J215729.43-074744.1, and, WINGS~J215733.47-074738.8, and they are
denoted by A, B, and, C, respectively, in figure~\ref{fig:a2399img} (b).

\section{X-ray spectra}\label{sec:a2399spectra}
To investigate the spatial distributions of the physical quantities that
characterize the cluster gas, we performed X-ray spectral analyses
using the XMM-Newton and Suzaku data.  We define two kinds of spectral-integration regions in subsection~\ref{subsec:spec_regions}, and we present 
the results of our analysis in subsections~\ref{subsec:spec_bridge} and
\ref{subsec:spec_azimuth}.

\subsection{Definitions of spectral regions} \label{subsec:spec_regions}
As figure~\ref{fig:a2399img} indicates, we extracted spectra from
the following two kinds of regions:
\begin{itemize}
\item[(i)] Bridge regions along the gas clumps (12 boxes; $1\arcmin.5\times 10\arcmin \times12 \simeq 100~{\rm kpc} \times 700~{\rm kpc}
  \times 12$).
\item[(ii)] Annular regions centered on the X-ray peak of the main clump after removal of point sources (section~\ref{sec:xray_image}). 
Considering the PSF size of the Suzaku XRT, five radial bins are chosen as follows: $0\arcmin < r < 2\arcmin$, $2\arcmin < r < 5\arcmin$, $5\arcmin < r < 8\arcmin$, $8\arcmin < r < 13\arcmin$,
  and $13\arcmin < r < 18\arcmin$, which correspond to $0 < r < 140$~kpc, $140 < r <
  340$~kpc, $340 < r < 540$~kpc, $540 < r < 880$~kpc, and $880 < r <
  1200$~kpc. Note that the first three rings are extracted from the `A2399 Center' field and thus cover a position angle $0-360\degree$, while the outer two rings in the east, north, and west directions are extracted from the `A2399 East', `A2399 North', and `A2399 West' fields of view, respectively and thus each region has a sector angle of about 90\degree.
  
\end{itemize}
For (i), we used only the XMM-Newton data to take advantage of its 
higher angular resolution and to avoid contamination from the point
sources (section~\ref{subsec:spec_bridge}). For (ii), we used the Suzaku data
to capitalize on its stable and low background characteristics, 
which enable us to study low surface brightness signals out to the
virial radius for the first time (section~\ref{subsec:spec_azimuth}).

\subsection{Bridge regions along the gas clumps} \label{subsec:spec_bridge}
To investigate the temperature and intensity profiles along the axis
connecting the main and sub gas clumps, we extracted XMM-Newton
spectra from the 12 boxes defined in (i) of the previous subsection (see
also figure~\ref{fig:a2399img} (b)).  We fitted the 0.5--11.0~keV spectra to 
the sum of the X-ray background, the cluster, and a soft proton background model, 
in the same manner as in \cite{2018PASJ...70...46B}.
%, ''apec$_{\rm LHB}$ +phabs * (apec$_{\rm GH}$ + powerlaw$_{\rm CXB})$ + apec$_{\rm ICM}$ +powerlaw$_{\rm SP}$''.  
For the X-ray background emission, we took into account the local
hot bubble (LHB), the Galactic halo (GH), and the cosmic X-ray
background (CXB) components. The parameters of
the X-ray background components are limited to the error ranges given
in table~\ref{tab:a2399background}.  The power-law function
representing the soft proton contamination was included with the
diagonal response files distributed by the XMM-Newton team.  The photon
indices for the MOS1 and MOS2 spectra are linked to each other,
as indicated in the guidelines given in the ESAS manual\footnote{https://heasarc.gsfc.nasa.gov/docs/xmm/esas/cookbook/xmm-esas.html}.
The Galactic column density $N_{\rm H}$ is fixed at 
$3.0 \times 10^{20}~{\rm cm}^{-2}$ based on the HI maps
\citep{2005A&A...440..775K}.  In the model for the hot cluster gas,
the redshift is fixed at 0.0579, and the metal abundance is fixed
at 0.22~solar, as obtained from the spectral analysis of the entire region.
 
\begin{figure}[!bt]
\begin{center}
\vspace{-0.cm}
  \includegraphics[width=100mm,angle=0]{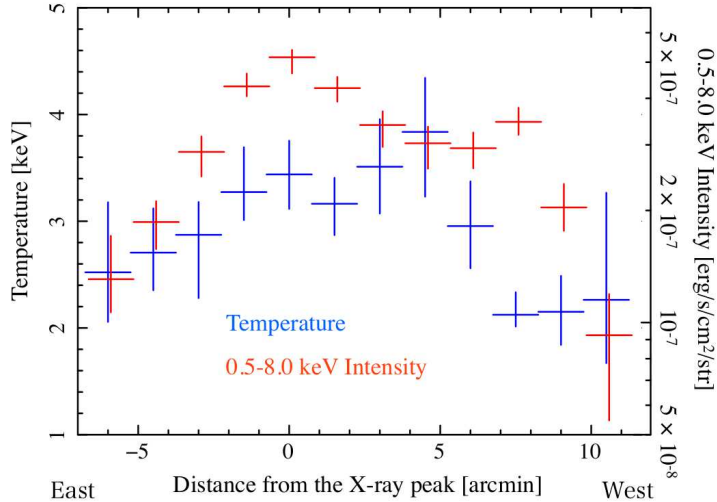}
\end{center}
\vspace{-0.cm}
\caption{Temperature (blue) and intensity (red) profiles across the
  main and sub gas clumps.  The horizontal axis shows the distance from
  the X-ray peak (negative is east, and positive is west; see figure \ref{fig:a2399img} (b)).  }
\label{fig:a2399tmpintensitymainsub}
\end{figure}

Figure~\ref{fig:a2399tmpintensitymainsub} shows the temperature and intensity profiles, and table~\ref{tab:spectraa2399brige} summarizes the data.  The intensity
profile shows two clear peaks, at $0\arcmin$ and $7\arcmin.5$
corresponding to the positions of the main and sub gas clumps.  In the
bridge region between the two gas clumps, the intracluster medium (ICM) shows a relatively flat temperature profile, with $kT\sim 3$~keV. The temperature
decreases to $\sim 2$~keV outside the bridge, and a steep gradient is
seen, particularly around the sub gas clump.  There are no prominent
temperature peaks at the positions of the X-ray-intensity peaks,
indicating a significant difference between the temperature and gas-density profiles.

%%%%%%%%%%%%%%%%%%%%%%%%%%%%%%%%%%%%%%%%%%
%%%%%%%%%%%%%%%%%%%%%%%%%%%%%%%%%%%%%%%%%%

%%%%%%%%%%%%%%%%%%%%%%%%%%%%%%%%%%%%%%%%%%
\begin{table}[htb]
\begin{center}
\caption{The best-fit parameters for the bridge region between the main and sub gas clumps}{ 
\hspace{0cm}
\begin{tabular}{lcccccc}
    \hline \noalign{\vskip2pt}
 Distance$^*$ &    $kT$       &   Norm$^\dagger$     &   $\chi^2/d.o.f.$  \\
    (arcmin)       &     (keV)     &         &                 \\  \hline
  -6.0       & $2.14_{-0.42}^{+0.71}$	&	$89_{-24}^{+25}$	&	300/323	\\
  -4.5       & $2.52_{-0.46}^{+0.66}$	&	$146_{-27}^{+45}$	&	363/358	\\
  -3.0       & $2.71_{-0.35}^{+0.42}$	&	$200_{-31}^{+27}$	&	437/440	\\
  -1.5       & $2.87_{-0.59}^{+0.31}$	&	$299_{-42}^{+30}$	&	574/541	\\
   0         & $3.27_{-0.26}^{+0.42}$	&	$421_{-25}^{+35}$	&	564/606	\\
   +1.5     & $3.44_{-0.32}^{+0.32}$	&	$492_{-47}^{+23}$	&	673/710	\\
   +3.0     & $3.16_{-0.29}^{+0.24}$	&	$424_{-34}^{+30}$	&	732/677	\\
   +4.5     & $3.51_{-0.44}^{+0.45}$	&	$321_{-41}^{+29}$	&	529/550	\\
   +6.0     & $3.84_{-0.61}^{+0.51}$	&	$276_{-40}^{+30}$	&	581/542	\\
   +7.5     & $ 2.95_{-0.40}^{+0.42}$	&	$302_{-35}^{+31}$	&	437/462	\\
   +9.0     & $2.12_{-0.11}^{+0.21}$	&	$419_{-32}^{+38}$	&	524/491	\\
   +10.5    & $2.15_{-0.31}^{+0.34}$	&	$245_{-33}^{+38}$	&	489/447	\\
    \hline
  \end{tabular}\label{tab:spectraa2399brige}
}
\end{center}
\begin{tabnote}
\footnotemark[$*$] The distance from the peak position in the main gas clump \\
\hspace{0.2cm}(minus: east, plus: west).\unskip\\
\footnotemark[$\dagger$] See the footnote, $*$, in table \ref{tab:a2399background}.\unskip\\
\end{tabnote}
\end{table}
%%%%%%%%%%%%%%%%%%%%%%%%%%%%%%%%%%%%%%%%%%

\subsection{Annular regions} \label{subsec:spec_azimuth}
To derive the three-dimensional structure of the temperature, density
and entropy of the hot gas, we de-projected the data, assuming a spherically symmetric gas distribution
(figure~\ref{fig:a2399img}a). We derived the distributions separately for each of the three
directions (east, north, and west) to study the spatial variations.  In
each direction, we fitted the five annular spectra in the 0.7--8~keV band simultaneously and the de-projection operation by
using the ``projct'' model in XSPEC. We assumed the APEC model for cluster
emission attenuated by Galactic absorption for each
annular bin.  We assumed the metal abundance to be constant, independent of radius, and fixed at 0.22~solar, as is also the case in section
\ref{subsec:spec_bridge}. Because the temperature was not well
constrained in the outer-most annular regions ($13\arcmin-18\arcmin$) in the
east and north directions, we linked the temperature of that region to the next-innermost annular region ($8\arcmin-13\arcmin$). The fit was
reasonable, giving $\chi^2$/degrees of freedom (d.o.f.)$= 408/382$, $484/414$, and 
$498/438$ for the east/north/west directions.

\begin{figure*}[htb]
\begin{center}
    \includegraphics[width=70mm,angle=0]{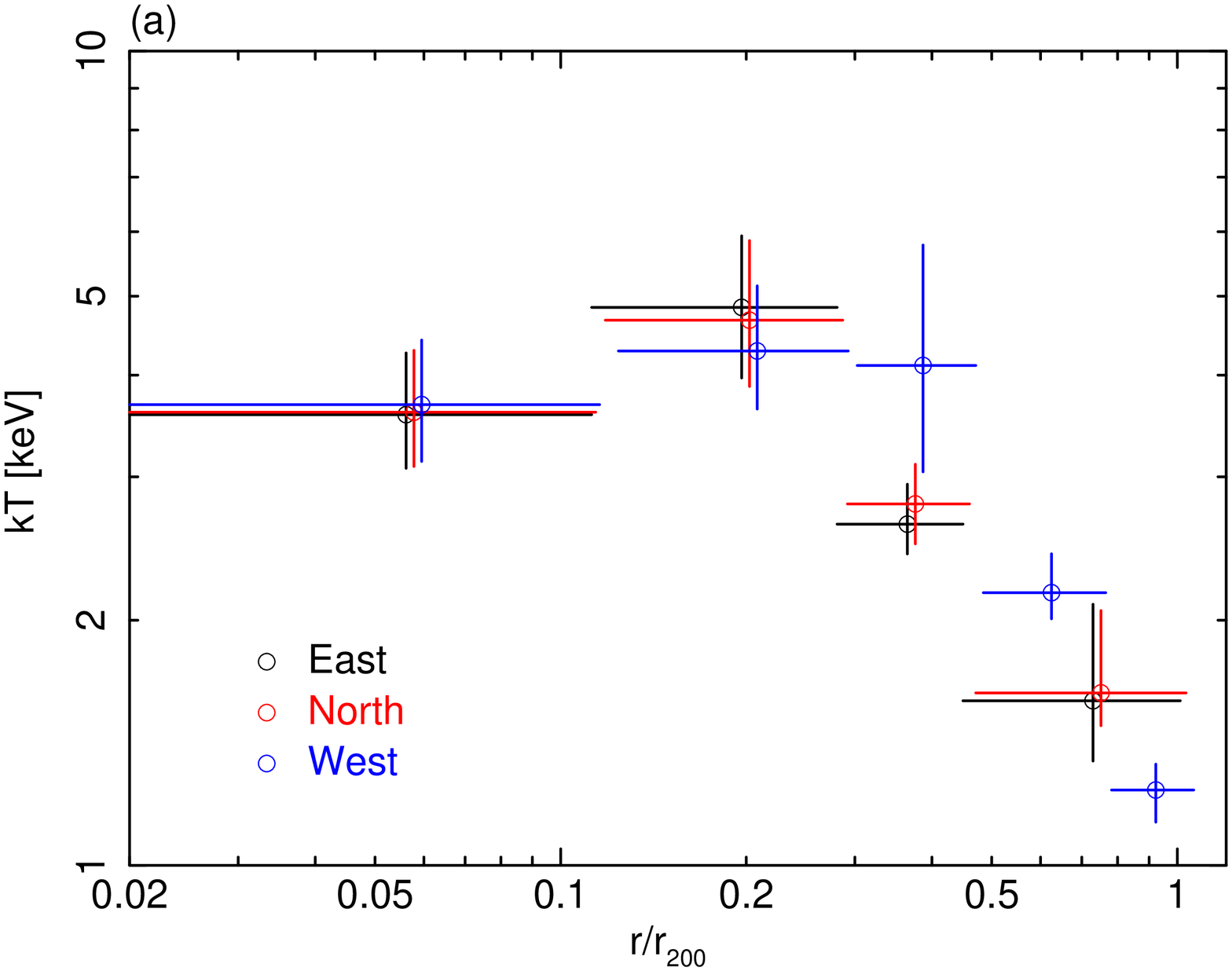}
    \includegraphics[width=70mm,angle=0]{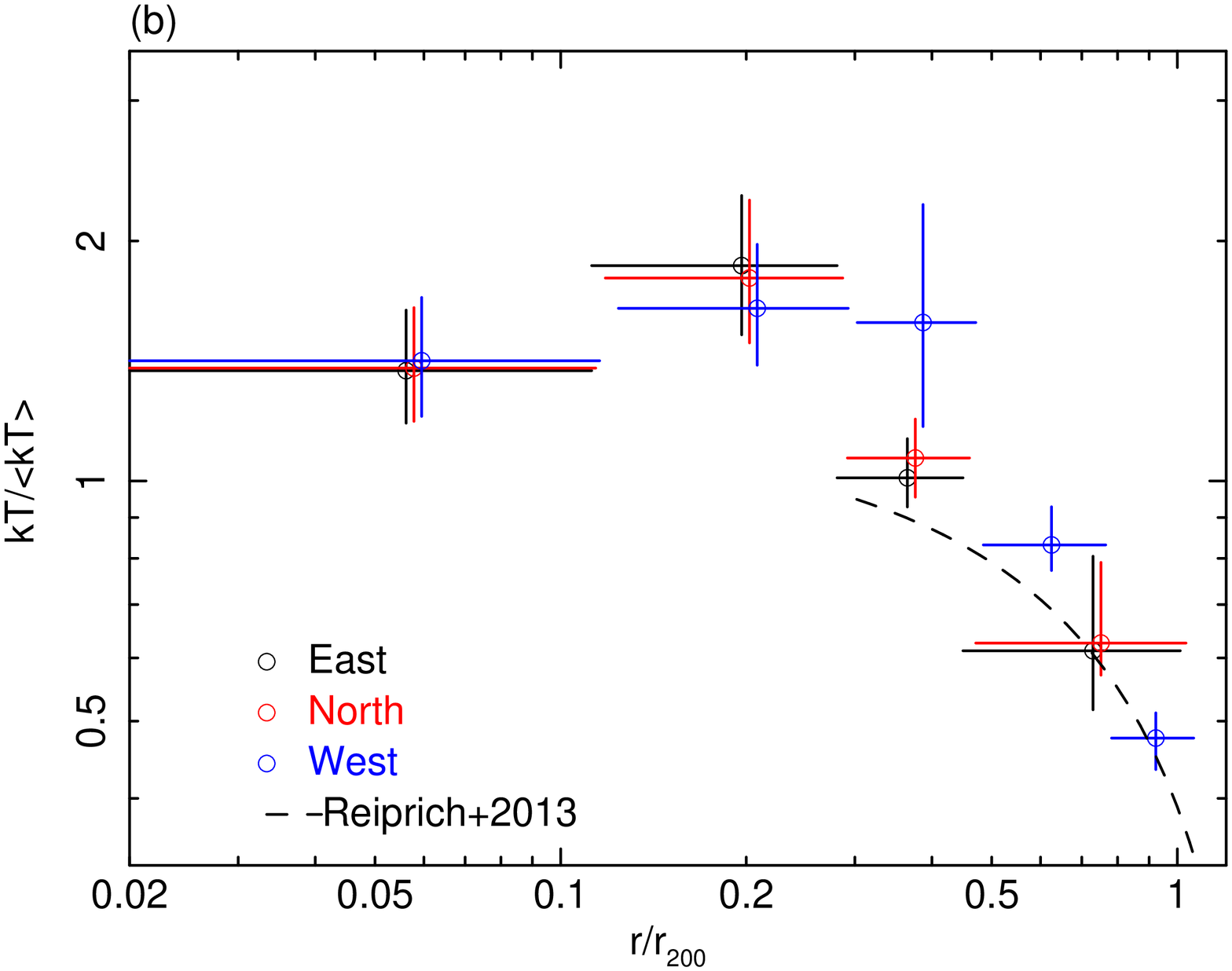}
    \includegraphics[width=70mm,angle=0]{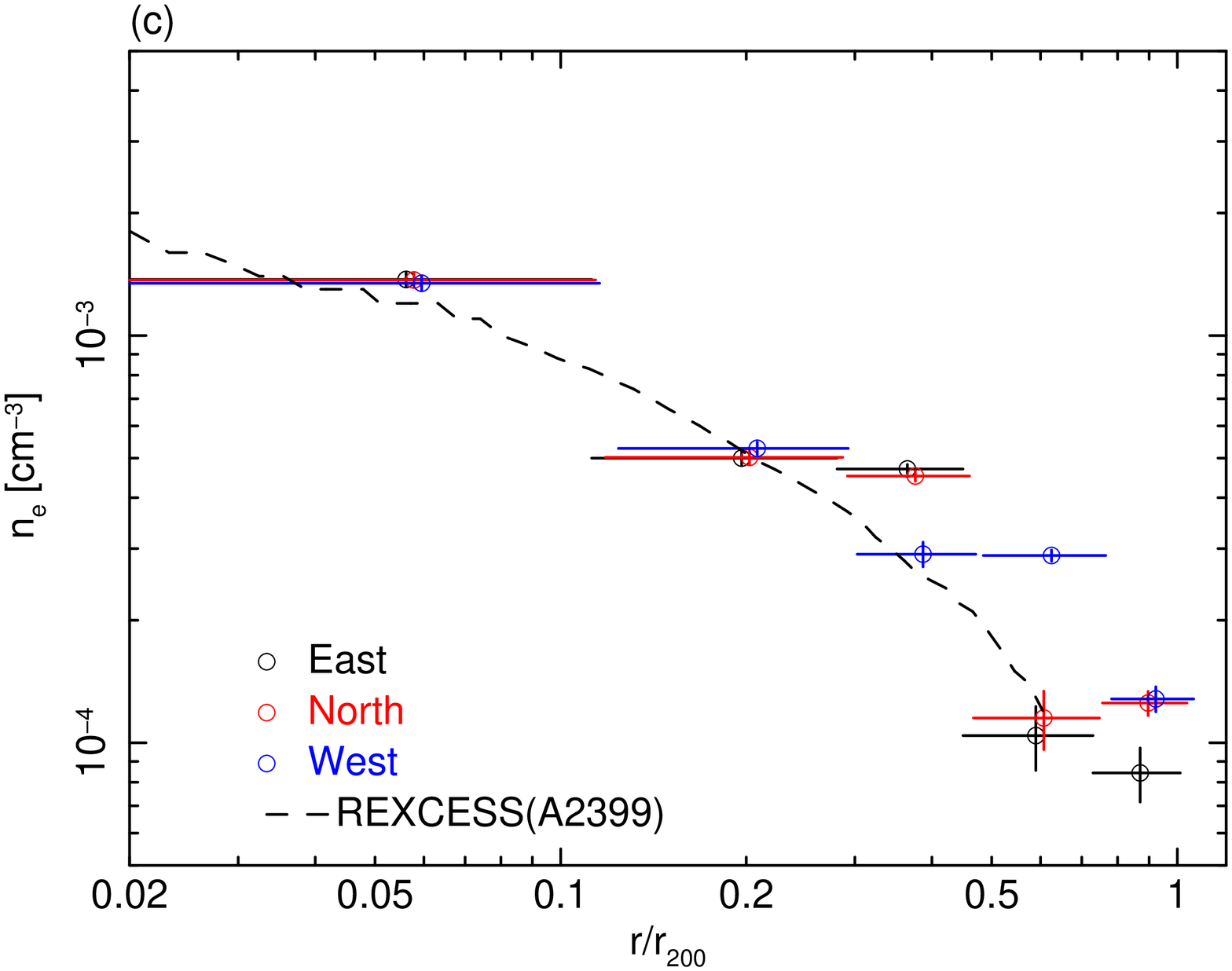}
    \includegraphics[width=70mm,angle=0]{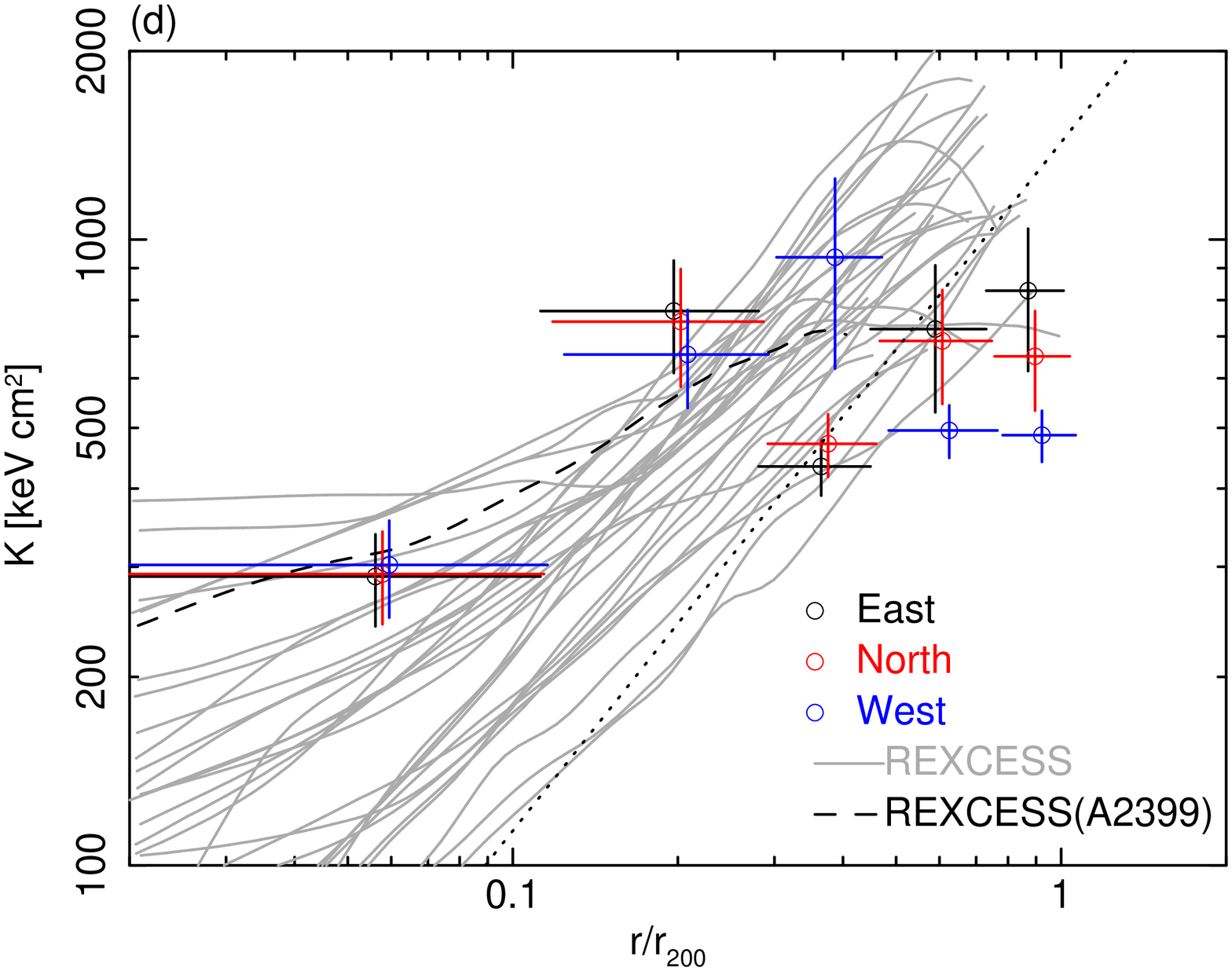}
  \end{center}
%\vspace{-0.0cm}
  \caption{(a) Temperature, (b) normalized temperature, (c) gas
    density, and (d) entropy profiles of A2399 derived from 
    de-projection of the Suzaku spectra.  The black, red, and
    blue symbols show the profiles in the east, north, and west
    directions, respectively.  The markers for the north (west)
    direction are shifted horizontally by 3 (6) \% for clarity.  In panel (b), the temperature is normalized relative to the mean value $\langle kT\rangle=2.6$~keV. The dashed line indicates the
   best-fit temperature profile obtained from the Suzaku
    observations of relaxed clusters \citep{2013SSRv..177..195R}.  In panel (c), the gas density derived by \cite{2008A&A...487..431C} is indicated by the dashed line. In panel (d), the entropy profiles of the REXCESS sample
    \citep{2010A&A...511A..85P} are also shown for comparison and the data for A2399 is denoted by the dashed line.  The
    dotted line indicates the baseline entropy profile
    \citep{2005MNRAS.364..909V} calculated for the mean temperature of
    A2399.  The radius is normalized by $r_{200}$.  }
\label{fig:a2399deproj}
\end{figure*}

Table~\ref{tab:deproj} lists the derived temperature values, and figure~\ref{fig:a2399deproj} (a) plots their profiles. 
The temperature is about 3.6~keV at the center and drops by a
factor of 2 in the outskirts.  In the ring $0.30<r/r_{200}<0.48$, the temperature in the west, $4.1^{+1.7}_{-1.1}$~keV, is statistically
consistent with the XMM result
(figure~\ref{fig:a2399tmpintensitymainsub}), and it is marginally higher
than in the other two directions.  

We calculated the electron density $n_e$ from the normalization of the APEC model. The central 
density is $1.4\times10^{-3}~{\rm cm^{-3}}$, which is extremely low in
comparison with other nearby clusters. Here the density is averaged within $0.1r_{200}$ because we used the radial bin-size of $2\arcmin$ considering the spatial resolution of Suzaku and assumed that the gas is uniform in each spherical shell. At a very small radius, however, the gas density can be higher; in fact, the previous XMM-Newton measurement yielded $3.5\times10^{-3}~{\rm cm^{-3}}$ at $r=4.7$~kpc or $0.004r_{200}$ \citep{2008A&A...487..431C}.
For A2399 West, a density excess is seen in the ring $0.48<r/r_{200}<0.76$, which can be attributed to
the western sub gas clump (figure~\ref{fig:a2399img} (b)).  Because of
the presence of this high-density clump, the de-projection analysis
leads to a relatively lower density bin in the ring $0.30<r/r_{200}<0.48$ in
comparison with the other directions, even though the surface brightness is apparently the same. 

We evaluated the entropy $K = kT n_e^{-2/3}$ from the temperature and gas density within $r_{200}$. The entropy
is very high; the azimuthally-averaged value is $\sim290 \pm 30~{\rm keV~cm^2}$ in the central $r<0.1r_{200}$ region, and flat
outside $0.1r_{200}$.  The entropy for $r/r_{200}>0.48$ in the west is
marginally lower than in the other directions, which again coincides with the location of the subclump. 
In comparison with the previous XMM-Newton study of the REXCESS sample, the Suzaku result for $r<0.36r_{200}$ agrees with their result within the measurement uncertainties. 
%Note that at a very small radius, the entropy is lower than the above value, i.e., $K=190~{\rm keV\,cm^{2}}$ for $r=0.01r_{200}$ \citep{2010A&A...511A..85P}.
For $r\sim0.36r_{200}$, there are significant azimuthal variations due to the clumpy structure and the difference in the entropy values can be attributed to the difference in their analysis methods because in the XMM-Newton deprojection analysis by \citet{2010A&A...511A..85P}, the western gas clump is excluded and the maximum radius is about a half of that used in the Suzaku analysis. 
 
%%%%%%%%%%%%%%%%%%%%%%%%%%%%%%%%%%%%%%%%%%
%%%%%%%%%%%%%%%%%%%%%%%%%%%%%%%%%%%%%%%%%%

%%%%%%%%%%%%%%%%%%%%%%%%%%%%%%%%%%%%%%%%%%
\begin{table*}[htb]
\begin{center}
\caption{Results of the de-projection analysis for A2399 East, North, and West}
\scalebox{0.9}[0.9]{
\begin{tabular}{llll|lll|lll}\hline\hline
            & \multicolumn{3}{c|}{A2399 East} &  \multicolumn{3}{c|}{A2399 North} &  \multicolumn{3}{c}{A2399 West} \\ \cline{2-4} \cline{5-7} \cline{8-10}
Region & $kT$ & $n_{e0}$ & $K$ & $kT$ & $n_{e0}$ & $K$ & $kT$ & $n_{e0}$ & $K$ \\ 
            & (keV) & ($10^{-4}~{\rm cm^{-3}}$) & (${\rm keV\,cm^{2}}$) & (keV) & ($10^{-4}~{\rm cm^{-3}}$) & (${\rm keV\,cm^{2}}$) & (keV) & ($10^{-4}~{\rm cm^{-3}}$) & (${\rm keV\,cm^{2}}$) \\ \hline
$0\arcmin-2\arcmin$	&	$	3.58 	_{	-0.50 	}^{+	0.68 	}	$	&	$	13.73 	\pm	0.53 	$	&	$	289 	\pm	49 	$	&	$	3.60 	_{	-0.51 	}^{+	0.69 	}	$	&	$	13.70 	\pm	0.53 	$	&	$	292 	\pm	49 	$	&	$	3.68 	_{	-0.55 	}^{+	0.74 	}	$	&	$	13.73 	\pm	0.54 	$	&	$	302 	\pm	53 	$	\\
$2\arcmin-5\arcmin$	&	$	4.84 	_{	-0.88 	}^{+	1.09 	}	$	&	$	5.00 	\pm	0.19 	$	&	$	769 	\pm	157 	$	&	$	4.67 	_{	-0.80 	}^{+	1.18 	}	$	&	$	5.02 	\pm	0.20 	$	&	$	739 	\pm	158 	$	&	$	4.28 	_{	-0.65 	}^{+	0.87 	}	$	&	$	5.29 	\pm	0.19 	$	&	$	655 	\pm	117 	$	\\
$5\arcmin-8\arcmin$	&	$	2.62 	_{	-0.21 	}^{+	0.31 	}	$	&	$	4.71 	\pm	0.12 	$	&	$	434 	\pm	44 	$	&	$	2.78 	_{	-0.30 	}^{+	0.33 	}	$	&	$	4.52 	\pm	0.14 	$	&	$	472 	\pm	54 	$	&	$	4.11 	_{	-1.07 	}^{+	1.67 	}	$	&	$	2.91 	\pm	0.20 	$	&	$	937 	\pm	315 	$	\\
$8\arcmin-13\arcmin$	&	$	1.59 	_{	-0.25 	}^{+	0.50 	}	$	&	$	1.04 	\pm	0.19 	$	&	$	719 	\pm	190 	$	&	$	1.63 	_{	-0.14 	}^{+	0.43 	}	$	&	$	1.15 	\pm	0.19 	$	&	$	688 	\pm	142 	$	&	$	2.16 	_{	-0.15 	}^{+	0.25 	}	$	&	$	2.88 	\pm	0.09 	$	&	$	495 	\pm	48 	$	\\
$13\arcmin-18\arcmin$	&	$\uparrow$						&	$	0.84 	\pm	0.13 	$	&	$	828 	\pm	212 	$	&	$\uparrow$						&	$	1.25 	\pm	0.09 	$	&	$	651 	\pm	118 	$	&	$	1.24 	_{	-0.11 	}^{+	0.09 	}	$	&	$	1.28 	\pm	0.09 	$	&	$	487 	\pm	46 	$	\\ \hline
\end{tabular}\label{tab:deproj}
}
\end{center}
\end{table*}
%%%%%%%%%%%%%%%%%%%%%%%%%%%%%%%%%%%%%%%%%%

\section{Discussion}\label{sec:discussion}
In this section, we discuss the thermodynamic properties
of A2399 based on the XMM-Newton and Suzaku results. Then, we compare
the X-ray and optical observations and discuss the dynamical state of the system.

%\subsection{Temperature, density, and entropy profiles}
\subsection{Thermodynamic properties of the cluster gas}
Previous X-ray observations have shown that the temperature profiles of
relaxed clusters exhibit similarities at large radii, once they are scaled by the
mean temperature and the virial radius
\citep[e.g.,][]{2005ApJ...628..655V,2007A&A...461...71P,2013SSRv..177..195R,2013MNRAS.432..554W,2013ApJ...766...90I,2018arXiv180500042G}. Figure~\ref{fig:a2399deproj} (b) compares the normalized temperature profile of A2399 and the
average of relaxed clusters observed with Suzaku
\citep{2013SSRv..177..195R}.  For $r>0.3r_{200}$, A2399 shows a decline similar to that of a relaxed cluster, while we found no significant cool-core structure for  $r<0.3r_{200}$.

\begin{figure}[hbt]
\begin{center}
\vspace{-0.cm}
  \includegraphics[width=100mm,angle=0]{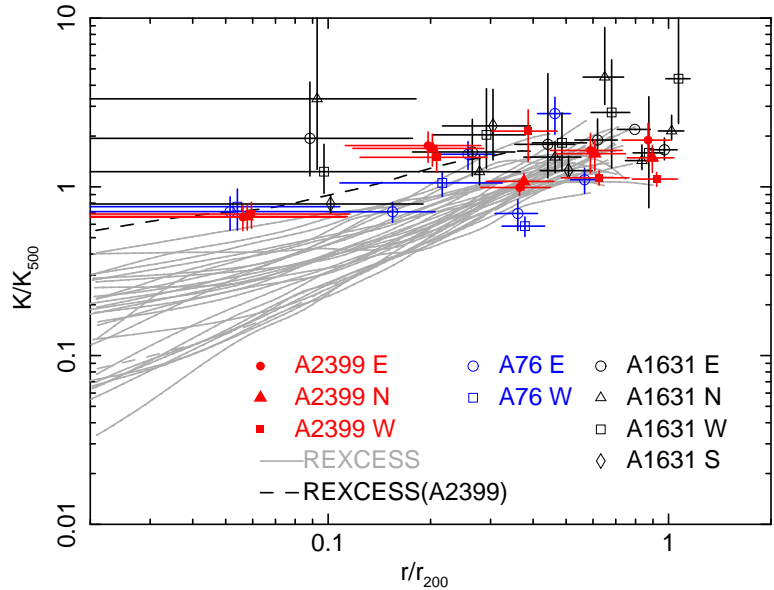}
\end{center}
\vspace{-0.cm}
\caption{Scaled entropy profile of A2399 in the east (red filled circle), north (red filled triangle), and west (red filled square) directions, as derived from the Suzaku XIS analysis. For comparison, the profiles of A76 in the east (blue open circle) and west (blue open square) and A1631 in the east (black open circle), north (black open triangle), west (black open square), and south (black open diamond) obtained from previous Suzaku observations are shown \citep{2013A&A...556A..21O, 2018PASJ...70...46B}.
The profiles of the REXCESS sample are shown by the gray lines, and A2399 (RXC~J2157-0747) is plotted by the dashed black line.}
\label{fig:scaled_entropy}
\end{figure}

Figure~\ref{fig:scaled_entropy} compares the A2399 scaled-entropy profile, as derived from our Suzaku analysis in section~\ref{subsec:spec_azimuth}, with those of the REXCESS sample derived from the XMM-Newton observations \citep{2010A&A...511A..85P}.  Here the Suzaku data of A2399 are scaled by the value of $K_{500}$ estimated from $M_{500}$ ($\sim$1.3 $h_{70}^{-1}$$\times$10$^{14}~$$M_{\odot}$) obtained by using the $M_{500}$--$Y_{{\rm X}}$ scaling relation in Table~1 and equation~(3) \citep{2010A&A...511A..85P}. In the central region, A2399 has the highest scaled entropy, $K/K_{500}\sim0.7$, in comparison with other REXCESS clusters as suggested by the previous study, and the present Suzaku result agrees well with that from XMM-Newton. We confirmed the difference in $M_{500}$ is only $\sim$20 \% corresponding to $\sim$10 \% in $K_{500}$ 
even though we adopt the $M_{500}$--T scaling relation \citep{2005A&A...441..893A} to obtain $M_{500}$.
At larger radii ($r\gtrsim0.3r_{200}$) up to around the virial radius, even in the east and north directions, the normalized entropy values seem to be systematically lower than 
those of relaxed clusters \citep{2012MNRAS.427L..45W} indicating that a large-scale event gives rise to the discrepancy in the whole cluster system.
%For $r\gtrsim0.3r_{200}$, there are significant azimuthal variations due to the clumpy structure. Note for $r\sim0.36r_{200}$, the difference in the scaled values can be attributed to the difference in their analysis methods because in the XMM-Newton deprojection analysis by \citet{2010A&A...511A..85P}, the western gas clump is excluded and the maximum radius is about a half of that used in the Suzaku analysis. 
The scaled-entropy profiles of other two LSB clusters, A76 and A1631, obtained from previous Suzaku studies \citep{2013A&A...556A..21O,2018PASJ...70...46B} are also shown in figure~\ref{fig:scaled_entropy}. Note that these two clusters are not a part of the REXCESS sample. The central scaled-entropy values of A2399 and A76 agree well with each other, while they tend to be lower than those of A1631. This mainly reflects a difference in electron density; A1631 has the lowest density of a few $10^{-4}~{\rm cm^{-3}}$ among known nearby clusters.
The uncertainty in the scaling factor may also affect the difference.

%%%%%%%%%%%%%%%%%%%%%%%%%%%%%%%%%%%%%%%%%%
%%%%%%%%%%%%%%%%%%%%%%%%%%%%%%%%%%%%%%%%%%

From numerical simulations that include gravitational heating
effect, \cite{2005MNRAS.364..909V} derived a `baseline' entropy
model for the cluster gas; $K~=~550~{\rm keV~cm^2}~(kT/1~{\rm
  keV})~(r/r_{200})^{1.1}$. Figure~\ref{fig:a2399deproj} (d) compares
the observed A2399 profile with the baseline model calculated for the
averaged temperature of $kT=2.6~{\rm keV}$. The azimuthally-averaged
central entropy, $294 \pm 29~{\rm keV~cm^2}$, is
significantly higher than that expected from gravitational heating alone ($\sim100~{\rm keV~cm^2}$ at $0.1 r_{200}$).

Non-gravitational heating processes can also contribute to the entropy
	excess, and numerical simulations suggest that AGN feedback plays a
major role \citep[e.g.,][]{2011MNRAS.417.1853D}.  Although no signs of
cavities or jets have been reported in A2399, we can assess the impact of AGN feedback as discussed in previous studies
\citep{2013A&A...556A..21O,2018PASJ...70...46B}.
% , since cavity- and jet-related feedbacks are suggested to be not generic \citep{2010RAA....10.1013W}.  
Using the $K$-band luminosity of the
brightest galaxy 2MASX~J21572939-0747443, log($L_K$/$L_{\rm K,
  \odot}$) = 11.5 \citep{2006AJ....131.1163S} and the empirical
relationship given by \cite{2010RAA....10.1013W}, we estimate the excess entropy in A2399 to be on the order of $1~{\rm keV\,cm^2}$. Therefore, we
conclude that the observed high entropy also cannot be explained by AGN feedback. Another likely origin of the excess entropy is disturbance by a cluster merger, which we will discuss in the following subsections.

\subsection{Comparison between X-ray and optical properties}\label{subsec:x-opt}
The galaxy distribution provides a clue as to the dynamical
state of a cluster.  For instance, if a cluster merger occurs,
substructures can be identified in both spatial
and velocity distributions of member galaxies \citep{1982PASP...94..421G}.  
Such substructures are detected in the adaptively
smoothed galaxy maps of many merging systems
\citep[e.g.,][]{2017MNRAS.468.1824O}.

The Omega WINGS survey \citep{2017A&A...599A..81M} has identified 234 spectroscopic galaxies as member galaxies of A2399.  
The smoothed galaxy map (see contours in figure~\ref{fig:a2399img}) 
shows several substructures, with  
the peak number density in the western and eastern galaxy clumps being 
$\sim90$ and $70~{\rm Mpc^{-2}}$, respectively.
Thus, the spatial offset between the gas and galaxy clumps is $>$100 kpc.
We have measured the central redshift and velocity dispersion 
of the galaxies that constitute the main and sub galaxy clumps.
There is no significant difference in central redshift  
between the galaxy clumps, while the observed velocity dispersion is 
$\sim800$ and $\sim600~{\rm km\,s^{-1}}$ in the main and sub galaxy clumps, respectively. The total mass is estimated to be $\sim 9$ and $\sim 3\times10^{14}~{\rm M_{\odot}}$, respectively \citep{2005ApJ...630..206F}.

%%%%%%%%%%%%%%%%%%%%%%%%%%%%%%%%%%%%%%%%%%
%\begin{figure*}
%\vspace{-0cm}
%\begin{center}
%\hspace{-0.0cm}
%   \includegraphics[width=160mm,angle=0]{./fig/x-ray_surface_brightness.pdf}
% \end{center}
%\vspace{-4.0cm}
%\caption{Left: exposure-corrected XMM-Newton image of A2399 in the
%  0.5--7.0 keV band with superposed contours of the galaxy distribution 
%  smoothed by a Gaussian kernel with FWHM = 200~kpc. Right: a close-up view
%  of the portion of the left figure around the jump structure.  The white solid boxes
%  are the extraction regions for the surface brightness measurements.  The yellow
%  dashed line indicates the boundary of the density jump.  The green solid
%  and dashed boxes are the extraction regions 
%  defined as the inner (A1) and the outer (A2) regions in our spectral analysis.  The white
%  arrow corresponds to the X-axis in figure
%  \ref{fig:surfacebrightness}.  The cyan circles identify point-source
%  regions excluded from the analysis. 
%*** 左のパネル BCGsの位置を表示
%  *** }
%\label{fig:surfacebrightnessimg}
%\end{figure*}
%%%%%%%%%%%%%%%%%%%%%%%%%%%%%%%%%%%%%%%%%%

\subsection{X-ray surface brightness discontinuity on the rim of the sub gas clump}
\begin{figure}[!hbt]
\begin{center}
   \includegraphics[width=100mm,angle=0]{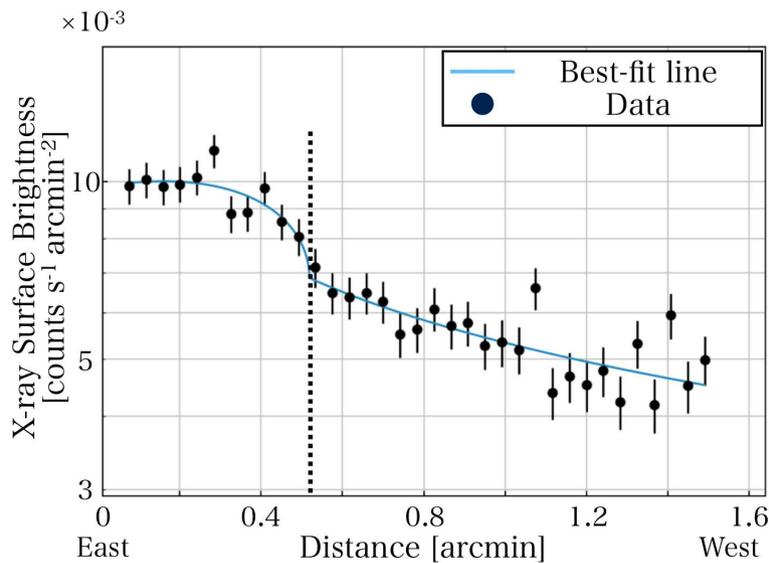}
 \end{center}
\caption{Surface brightness profile around the rim of the sub gas clump as shown in the white rectangular region in
  figure~\ref{fig:a2399img} (c).  The X-axis labeled ``Distance'' is defined from the east to the west direction in figure~\ref{fig:a2399img}.  
  The blue line indicates the
  best-fit surface-brightness profile model that is calculated from
  the double power-law model (equation~\ref{eq:double_pl}) The error
  bars indicate the 1$\sigma$ statistical uncertainties. The dotted line corresponds to the position of the discontinuity.}
\label{fig:surfacebrightness}
\end{figure}

To constrain the dynamical state of the system, we searched for
discontinuities in the surface brightness distribution.  As seen from
figures \ref{fig:a2399img} (c) and
\ref{fig:surfacebrightness}, there is a clear surface-brightness
discontinuity near the rim of the western sub gas clump. Therefore we modeled the
surface brightness by integrating the following double power-law model
for the electron-density profile along the line of sight
\citep[e.g.,][]{2009ApJ...704.1349O}:
\begin{equation}
n_{\rm e} (r) = 
%\begin{cases}
%  \begin{array}{ll}
    \ n_0 \left(\frac{r}{r_s}\right)^{-\alpha1}   (r < r_s), \\\\
    \frac{n_0}{j_{12}} \left(\frac{r}{r_s}\right)^{-\alpha2}   (r \geq r_s). \label{eq:double_pl}
%  \end{array}
%\right
%　\end{cases}
\end{equation}
Here we assume spherically symmetry for simplicity, and $n_0$, $r_s$, 
and $j_{12}$ represent the central density, the radius at the
point of discontinuity, and the density-jump parameter, respectively.
The quantities $\alpha1$ and $\alpha2$ are the power-law indices for the high-density
and low-density regions, respectively.

%%%%%%%%%%%%%%%%%%%%%%%%%%%%%%%%%%%%%%%%%%
%%%%%%%%%%%%%%%%%%%%%%%%%%%%%%%%%%%%%%%%%%
%  the   three dimensional electron density profile with the density discrepancy.  

A case with no density jump, that is, the model with the jump parameter
fixed at 1, does not reproduce the observed profile well. However, when the jump parameter is set free, the fit improves significantly (the null hypothesis
probability is 1.2\%, according to the F-test). The best-fit parameter is $j_{12}
= 1.6 \pm 0.2$.

The XMM spectral analysis demonstrated that the temperatures of the
upstream and downstream regions 
across the plane of the discontinuity are 1.83$^{+0.26}_{-0.17}$ keV and
3.71$^{+1.27}_{-1.04}$ keV, respectively.
We find that pressure equilibrium holds at the boundary of the
density jump to within the error bars. Thus the density discontinuity 
originates from a cold front rather than a shock.

\subsection{A possible scenario}
As is the case for A1631 \citep{2018PASJ...70...46B}, 
the observational characteristics in both X-ray and optical-- 
i.e., no prominent cool-core structure, a flatter density profile, 
excess entropy at the center, and a spatial offset between the peaks 
in the X-ray surface brightness and galaxy number density distributions-- 
suggest that the A2399 system is also in some stage of merging.

By utilizing hydrodynamic simulations, \cite{2011ApJ...728...54Z}
has studied the effect of ICM mixing as a function of a mass ratio and 
impact parameter of a cluster collision. 
The simulations reproduce a high-entropy core--where the
initial core is replaced by a low-density, high-temperature and
high-entropy core--and in which the temperature and gas-density profiles become
flatter, relative to those of the initial profiles. 
Therefore, we consider that the excess entropy in A2399 can be attributed to 
ICM mixing during a cluster merger. 

Underluminosity in X-rays at the given velocity dispersion $\sigma = 729 \pm 35~{\rm km~s^{-1}}$ \citep{2017A&A...599A..81M} extracted by using the whole member galaxies is also seen, with the observed luminosity being approximately 1/3 that expected from the $L-\sigma$ relation \citep{2011A&A...526A.105Z} predicted from the hydrodynamic simulations \citep[e.g.,][]{2011ApJ...728...54Z}.
\citet{2015Ap.....58..328T} argues that some merging clusters show a higher ($>$1.2) or smaller ($<$0.8) ratio of n$_{l}$/n$_{h}$, where n$_{l}$ and n$_{h}$ correspond to the numbers of galaxies with velocities lower and higher than the mean velocity of all member galaxies, and the observed ratio for A2399 is 0.7$\pm$0.1. 
In the case of A2399, the significant substructures seen in both X-ray and optical data and the discrepancy in the X-ray peak positions between the temperature and 
surface brightness also support the merger scenario \citep[e.g.,][]{2004ApJ...606..819M,2015Sci...347.1462H}.

Furthermore, the main gas clump seen in X-rays is located 
between the main and sub galaxy clumps observed in the optical, and the cold front is found around the rim of the gas clump. The X-ray morphology is elongated along an axis connecting the galaxy clumps, and there is little difference in redshift between the two galaxy clumps (see section~\ref{subsec:x-opt}), indicating that the cluster-cluster collision occurred nearly in the plane of the sky.
Thus, we conclude that A2399 may be a system in a head-on collision, 
like the Bullet cluster　\citep[e.g.,][]{2007PhR...443....1M}.
Deep X-ray observations and weak lensing will enable the determination of further details, such as shock structures and dark matter distributions.
Because this LSB cluster resides in a supercluster,  
the merger events may have resulted from its dense environment.

\section{Summary}
To study the nature of low surface brightness
clusters, we have analyzed the XMM-Newton and Suzaku data of an
LSB cluster, A2399 at $z=0.058$.  The X-ray image is
elongated and has a bimodal structure, while the temperature peak is
significantly offset from the X-ray emission peaks. We have derived the profiles of
gas temperature, density, and entropy out to the viral
radius. The observed high entropy in the central region, $\sim 300~{\rm keV\,cm^2}$, 
is comparable to other LSB clusters like A76 and A1631, and the scaled entropy of A2399 is the highest among the entire REXCESS sample. The galaxy map
also exhibits main and sub clumps; however there is a considerable offset
between the peak positions in the main galaxy clump and the main
gas clump. Furthermore, we have detected the brightness and temperature
discontinuities near the sub galaxy clump in the west, providing evidence for the presence of a cold front. Therefore, our results
support the scenario that this cluster has experienced a merger in the
plane of the sky, 
%, as is also the case for the Bullet cluster
and the excess central entropy and complex X-ray and
optical morphologies are caused by the merger. Since the X-ray
properties of LSB clusters tend to deviate from the scaling
relations of known nearby clusters, we expect the present study to have a
significant impact on the detection techniques and interpretations of faint,
diffuse clusters in future cluster surveys.

\section*{Acknowledgement}
We thank all the Suzaku team members for their support.  IM is
grateful to Y. Tawara for the useful advice in the analysis and thanks
Y. Fujita for his insightful comments. This work was supported in part
by JSPS KAKENHI grant number 26220703 (IM) and 16K05295 (NO). HB and GC acknowledge
support from the DFG Transregio Program TR33 and the Munich Excellence
Cluster ``Structure and Evolution of the Universe''.  GWP acknowledges
funding from the European Research Council under the European Union's
Seventh Framework Programme (FP7/2007−2013)/ERC grant agreement
No. 340519.

%\bibliography{reference}

\end{document}